\documentclass[conference]{IEEEtran}
\IEEEoverridecommandlockouts
\usepackage{cite}
\usepackage{amsmath,amssymb,amsfonts}
\usepackage{algorithmic}
\usepackage{graphicx}
\usepackage{textcomp}
\usepackage{xcolor}

\usepackage{subcaption}
\usepackage{caption}
\usepackage{booktabs}
\usepackage{colortbl}

\def\BibTeX{{\rm B\kern-.05em{\sc i\kern-.025em b}\kern-.08em
    T\kern-.1667em\lower.7ex\hbox{E}\kern-.125emX}}
\begin{document}

\title{Consistency of Muscle Synergies Extracted via Higher-Order Tensor Decomposition Towards Myoelectric Control\\
\thanks{{*} Corresponding author;  email: {\tt\small ahmed.ebied@ed.ac.uk}}
}

\author{\IEEEauthorblockN{ Ahmed  Ebied{*}}
\IEEEauthorblockA{\textit{School of Engineering}\\
\textit{Institute for Digital Communications} \\
\textit{University of Edinburgh}\\
Edinburgh, United Kingdom \\
}
\and
\IEEEauthorblockN{Eli Kinney-Lang}
\IEEEauthorblockA{\textit{School of Engineering}\\
\textit{Institute for Digital Communications} \\
\textit{University of Edinburgh}\\
Edinburgh, United Kingdom \\
}
\and
\IEEEauthorblockN{Javier Escudero}
\IEEEauthorblockA{\textit{School of Engineering}\\
\textit{Institute for Digital Communications} \\
\textit{University of Edinburgh}\\
Edinburgh, United Kingdom \\}
}

\maketitle

\begin{abstract}

In recent years, muscle synergies have been proposed for proportional myoelectric control. Synergies were extracted using matrix factorisation techniques (mainly non-negative matrix factorisation, NMF), which requires identification of synergies to tasks or movements. In addition, NMF methods were viable only with a task dimension of 2 degrees of freedoms (DoFs). Here, the potential use of a higher-order tensor model for myoelectric control is explored. We assess the ability of a constrained Tucker tensor decomposition to estimate consistent synergies when the task dimensionality is increased up to 3-DoFs. Synergies extracted from $3^{rd}$-order tensor of 1 and 3 DoFs were compared. Results showed that muscle synergies extracted via constrained Tucker decomposition were consistent with the increase of task-dimension. Hence, these results support the consideration of proportional 3-DoF myoelectric control based on tensor decompositions.

\end{abstract}

\begin{IEEEkeywords}
Muscle synergy, Myoelectric control, Tensor factorisation, Tucker decomposition, EMG
\end{IEEEkeywords}

\section{Introduction}

Recently, the concept of modular organisation of motor tasks and muscle synergy has been investigated extensively and it has been accepted as a framework to understand motor control \cite{DAvella2015}. The notion of modularity arises from the complexity and redundancy of motor control \cite{DAvella2003}. Thus, the muscle synergy concept posits that a  motor control task is performed using a combination of a few synergies rather than controlling individual muscles. Despite the debate about the neural origin of muscle synergies \cite{TRESCH}, they have been useful in clinical \cite{Pons2016a} and biomechanical studies \cite{Nazifi2017}. Moreover, synergies have been utilised in myoelectric control through classification \cite{Ma2015a,Rasool2016} or proportional strategies \cite{Jiang2014b,Ison2015,Lin2017}.

According to the time invariant model of muscle synergies \cite{Tresch1999}, the electrical muscle activity, recorded by electromyography (EMG), is defined by a set of synchronised synergies weighted by time-varying functions. As a result, the identification of muscle synergies from multi-channel EMG signals is a blind source separation problem. Several matrix factorisation methods have been proposed to extract muscle synergies by treating the multi-channel EMG as a matrix with modes (dimensions) channels $\times$ time. Non-negative matrix factorisation (NMF) \cite{Lee1999} has been the most prominent technique for synergy extraction \cite{Tresch2006,Ebied2018}. However, matrix factorisation methods have limitations. For instance, in biomechanical studies, identifying shared muscle synergies requires to apply NMF repetitively on each task and/or subject, then relying on metrics such as the correlation coefficient to identify the shared and task-specific synergies. This makes such approach complex and unreliable \cite{Ebied2018b}. Moreover, the performance of proportional myoelectric control based on NMF synergies degrades significantly with the of increase task-space dimension into 3 degree of freedoms (DoF) of movement \cite{Jiang2014b,Ma2015a}. In addition, the current approaches assign two synergies for each DoF (1 synergy per task). Thus, the number of synergies needed for control increases with the number of tasks \cite{DeRugy2013}.

Multi-channel EMG data tend to be represented in the form of matrix -- $2^{nd}$-order data array -- with spatial and temporal modes. However, in most EMG studies, data are naturally structured in higher-order form, such as repetitions of subjects and/or movements. Therefore, the muscle activity naturally fits into a higher-order tensor model. We have recently introduced a higher-order tensor model for muscle synergy extraction \cite{Ebied2017}, where EMG data were organised in a higher-order tensor form ($3^{rd}$ or $4^{th}$ order) rather than matrix. In the higher-order tensor model, muscle synergies are estimated via tensor decomposition methods, which provide several advantages over matrix factorisation such as compactness, uniqueness of decomposition, and generality of the identified components \cite{Cichocki2014}. The most common tensor decomposition methods are Tucker \cite{Tucker1966a} and Parallel Factor Analysis \cite{Harshman1970}.We have discussed the use of both methods for muscle synergy extraction from $3^{rd}$-order tensors with modes (channels $\times$ time $\times$ repetitions). The tensors were constructed by stacking repetitions of multi-channel EMG for two tasks from a single degree of freedom (DoF). In addition, a constrained Tucker method was developed to identify task-specific and shared synergies across each DoF in a direct less complex way compared to NMF \cite{Ebied2018b}. 

Here, the potential use of tensor synergies in proportional myoelectric control is explored by first analysing their consistency. We hypothesise that, in order for the tensor synergies to be useful in proportional myoelectric control, they need to be consistent when extracted from different numbers of DoF. Thus, we investigate the ability of a constrained Tucker decomposition to extract muscle synergies from 3-DoFs (6 tasks) tensors. The $3^{rd}$-order tensors will be constructed by repetitions of the 6 tasks (movements) that forms the main 3 wrist's DoFs. Synergies estimated from this 3-DoFs tensors will be compared with synergies identified by decomposition of 1-DoF tensors to test if they are consistent when increasing the task dimensionality from 1 to 3 DoFs. 

\section{Methods}

\subsection{Data and tensor construction}
\label{sec:DataTensor}
In this study, six tasks, or movements, were selected from the publicly available Ninapro first data-set \cite{Atzori2015a} which consists of 53 hand, wrist, and finger movements in total. The wrist motion and its three DoFs: wrist flexion and extension (DoF1), wrist radial and ulnar deviation (DoF2); and wrist supination and pronation (DoF3); are investigated since they are essential for myoelectric control \cite{Jiang2014b}. Each task has 10 repetitions from 27 able-bodied subjects recorded by a MyoBock 13E200-50 system (Otto Bock HealthCare GmbH). The data-set includes 10-channel surface EMG signals rectified by root mean square and sampled at 100Hz.

Tensors were constructed by stacking repetitions of the 10-channel EMG segments to form  $3^{rd}$-order tensors with modes (channels $\times$ time $\times$ repetitions). Two types of tensors were used in this study. The first is a larger 3-DoFs tensor which consists of repetitions from the six wrist tasks stacked together. On the other hand, a smaller tensor for single DoF is created from repetitions of the 2 tasks of that DoF.

\subsection{Constrained Tucker decomposition.}
\label{sec:constTucker}
Higher-order tensors can be decomposed into their main components in a similar way to  matrix factorisation. Tucker decomposition \cite{Tucker1966a} is one of the most prominent models for tensor factorisation. In a Tucker model, the $3^{rd}$-order tensor $\underline{\mathbf{X}}\in \mathbb{R}^{i_{1}\times i_{2}\times i_{3}}$ is decomposed into a smaller core tensor ($\underline{\mathbf{G}}\in \mathbb{R}^{j_{1}\times j_{2}\times j_{3}} $) transformed by a matrix across each mode (dimension) \cite{Kolda2008c}, where the core tensor determine the interaction between those matrices as the following:
\begin{equation}\label{eq_tucker_model}
\underline{\mathbf{X}} \approx \underline{\mathbf{G}} \times_{1}\mathbf{B}^{(1)} \times_{2}\mathbf{B}^{(2)} \times_{3}\mathbf{B}^{(3)} 
\end{equation}
where  $ \mathbf{B}^{(n)}\in \mathbb{R}^{i_{n}\times j_{n}}$ are the components matrices transformed across each mode while ``$\times_{n}$" is  multiplication  across the $n^{th}$-mode \cite{Kolda2008c}. 

The estimation of core tensor and  component matrices according to the Tucker model is typically carried out using the Alternating Least Squares algorithm (ALS). ALS has two main phases. The first one is initialisation of components and core tensor. The second phase is a series of iterations to minimise the loss function between the original data and its model \cite{Smilde2004}. The least squares loss function across the first mode is:
\begin{equation}\label{eq_loss_fun_Tucker}
argmin_{\mathbf{B}^{(1)},\mathbf{B}^{(2)},\mathbf{B}^{(3)},\mathbf{\underline{G}}} \|\mathbf{\underline{X}-\mathbf{B}^{(1)}\underline{\mathbf{G}}(\mathbf{B}^{(3)}\otimes \mathbf{B}^{(2)})^{T}} \|^{2}
\end{equation}
where $\otimes$ is Khatri-Rao product which is the column-wise Kronecker product.  This function is solved  by fixing the two factors from ($\mathbf{B}^{(1)}, \mathbf{B}^{(2)}, \mathbf{B}^{(3)}$) and computing the third unfixed factor alternatively.. The main drawback of ALS is that it cannot guarantee convergence to a stationary point \cite{Comon2009}. Hence, multiple constraints on the initialisation and iteration phases are needed to improve the estimation and achieve uniqueness of the solution. Moreover, the constrained Tucker model may help achieve interpretable results that do not contradict prior knowledge \cite{Cichocki2014}. 

In this study, a constrained Tucker decomposition was utilised to extract muscle synergies. The number of components were designed to be the same as number of tasks ($n_{tsk}$) for the time and repetition modes. On the other hand, the number of spatial components (synergies) would be the sum of the number of task-specific and shared synergies. That is, $1.5 n_{tsk}$ since we assume one shared synergy for each DoF (2 tasks). Three constraints have been imposed on this $[n_{tsk}, 1.5 n_{tsk}, n_{tsk}]$ Tucker model. Two of them were used during the initialisation phase and one constraint was implemented in the iteration phase of ALS. Both core tensor and repetition mode were initialised and fixed to identify the spatial mode components. Since each component is linked to one task, each component in this mode was designed to have a value of $1$ for a repetition of the considered movement and $0$ otherwise. The core tensor is initialised into a value of $1$ between each component in the (time$\backslash$repetition) modes and its respective spatial synergy (either task-specific or shared) and $0$ otherwise. The task specific synergies are linked to one components (time$\backslash$repetition) modes while the shared synergies are linked to the two components that form the desired DoF. For the iteration phase, non-negativity constraint have been imposed  on temporal and spatial modes because of the additive nature of synergies \cite{Ebied2018}.

\subsection{Comparison between single and 3-DoFs tensors synergies}

\begin{figure*}[t!]
    \centering
\begin{subfigure}{0.9\textwidth}
\includegraphics[width=\textwidth]{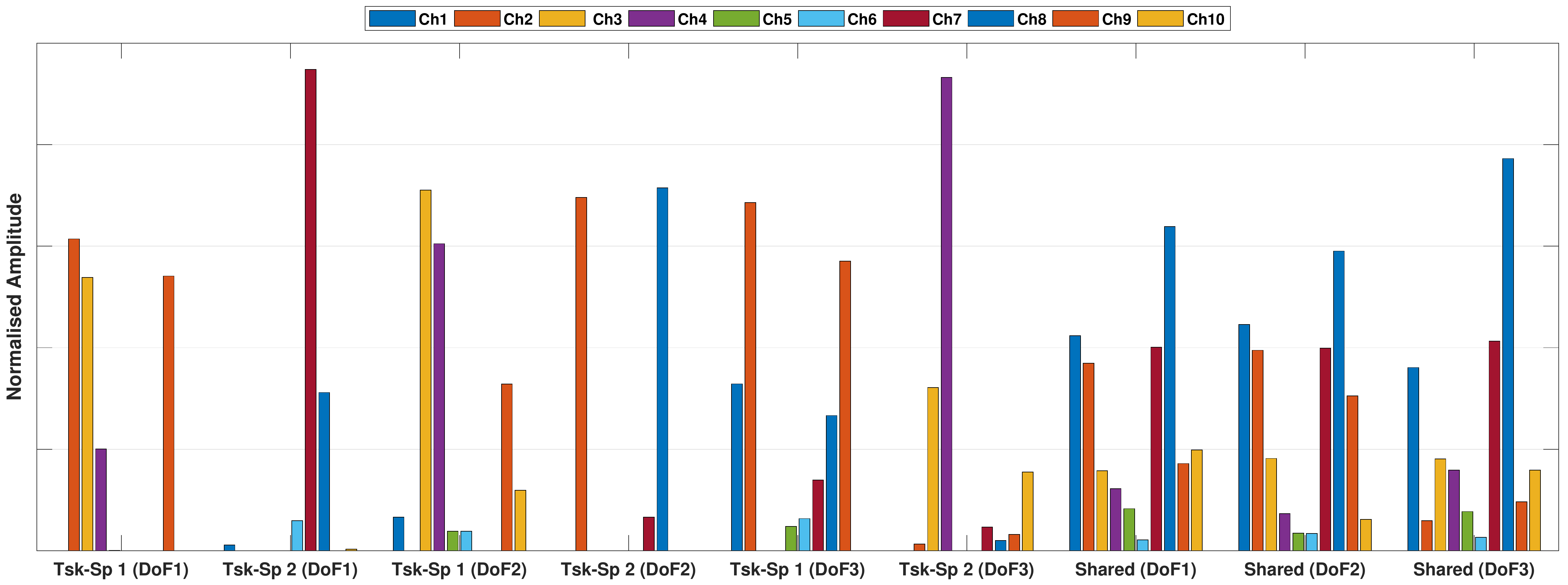} 
\caption{\textit{Synergies extracted from a tensor with all 3 DoFs.}}
\label{fig:Synergy3DoFs}
\end{subfigure}

\begin{subfigure}[b]{0.32\textwidth}
\centering
\includegraphics[width=\textwidth]{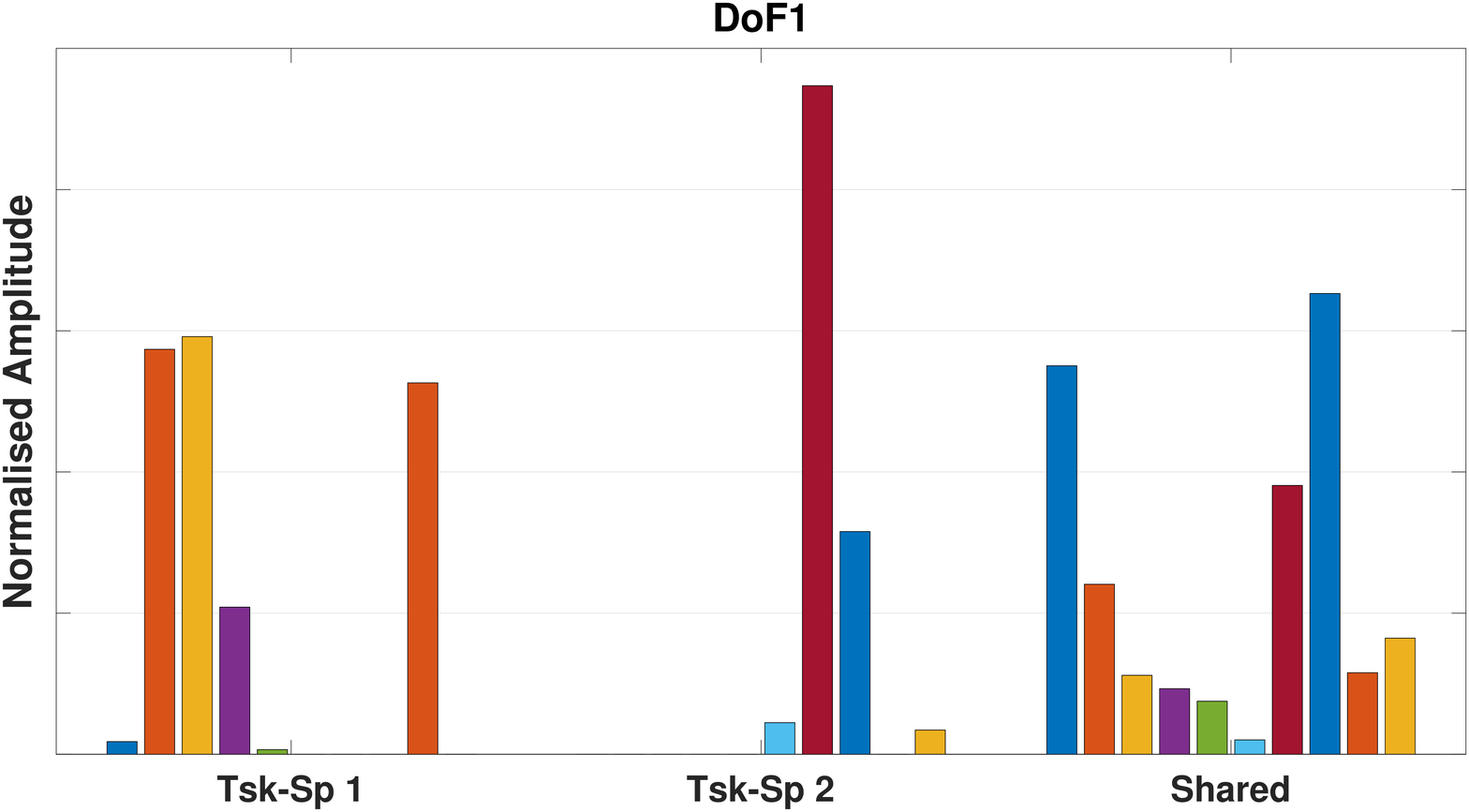}  
\caption{\textit{Synergies extracted from the DoF1 tensor.}}
\label{fig:SynDoF1}
\end{subfigure}
\begin{subfigure}[b]{0.32\textwidth}
\centering
\includegraphics[width=\textwidth]{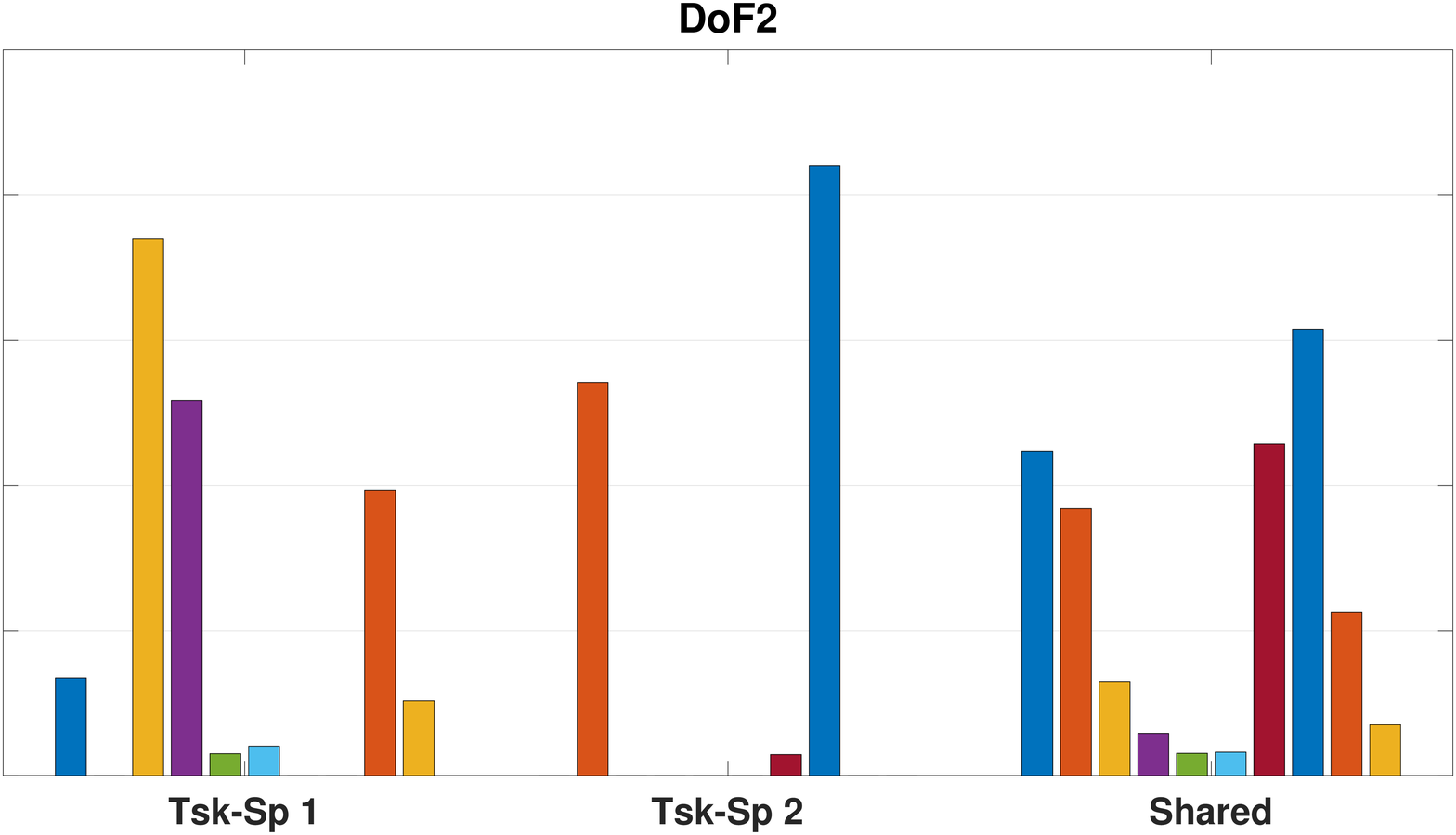} 
\caption{\textit{Synergies extracted from the DoF2 tensor.}}
\label{fig:SynDoF2}
\end{subfigure}
\begin{subfigure}[b]{0.32\textwidth}
\centering
\includegraphics[width=\textwidth]{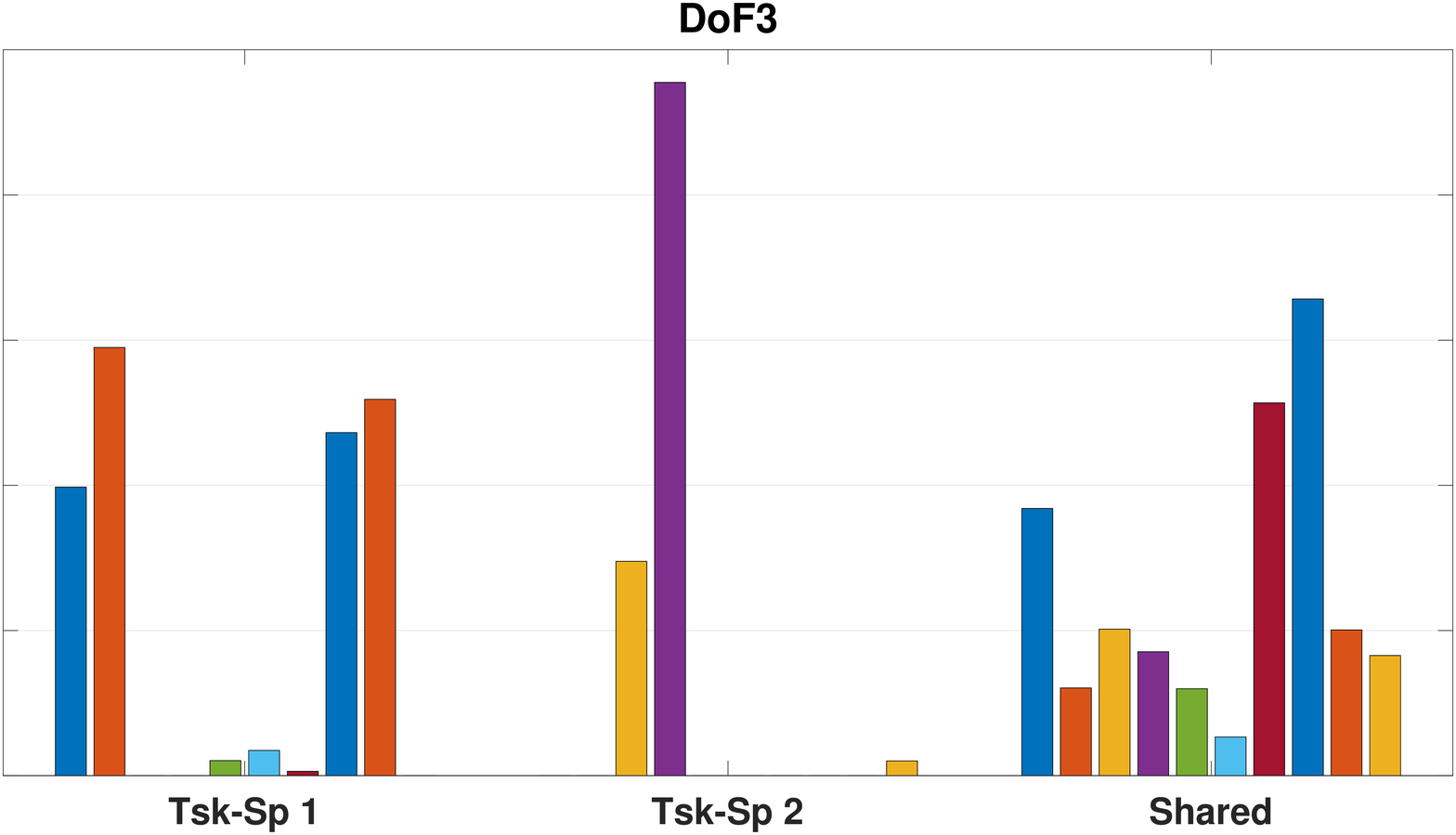} 
\caption{\textit{Synergies extracted from the DoF3 tensor.}}
\label{fig:SynDoF3}
\end{subfigure}
    \caption{The spatial mode (synergies) estimated via constrained Tucker method from a $3^{rd}$-order tensor of all three wrist DoFs (Panel \ref{fig:Synergy3DoFs}) and synergies estimated separately  from  $3^{rd}$-order tensor of DoF1 (Panel \ref{fig:SynDoF1}), DoF2 (Panel \ref{fig:SynDoF2}) and DoF3 (Panel \ref{fig:SynDoF3}) using the same constrained Tucker method for subject 1.}
    \label{fig:SynergyExample}
\end{figure*}

We apply the constrained Tucker model discussed in \ref{sec:constTucker} on both single (1-DoF) and 3-DoFs $3^{rd}$-order tensors described in \ref{sec:DataTensor}. The 1-DoF tensor is decomposed into [2,3,2] components giving 3 muscle synergies, 2 of which are task-specific and 1 is shared across the DoF. On the other hand, the 3-DoFs tensor is decomposed into [6,9,6] components, with 9 synergies (6 task-specific and 3 shared).

For each subject, synergies are extracted from both tensors, then compared against each other to test the consistency of the estimated synergies with the increase of task-dimensionality from 1-DoF to 3-DoFs.Two similarity indices (Pearson Correlation coefficient and cosines of angles) were computed between each synergy estimated from single DoF tensors and its respective synergy from the 3-DoF tensor for the 27 subjects. The mean values for correlation coefficients of synergies are calculated across subjects.

\section{Results}

Muscle synergies of the three wrist DoFs were extracted using constrained Tucker decomposition applied on two $3^{rd}$-order tensors setup.  The first one is a 3-DoF  tensor including the repetition of the all 6 tasks decomposed to [6,9,6] components with 9 synergies (6 task-specific and 3 shared) as shown in Fig.~\ref{fig:Synergy3DoFs}. The other approach uses a 1-DoF tensor including the repetition of 2 tasks decomposed by constrained Tucker decomposition to [2,3,2] components where 2 task-specific synergies and 1 shared are identified. This is done for each DoF separately as shown in Fig.~\ref{fig:SynDoF1}, \ref{fig:SynDoF2} and \ref{fig:SynDoF3} for DoFs 1,2 and 3 respectively.

\begin{figure*}[t!]

  \begin{subfigure}[b]{0.32\textwidth}
    \includegraphics[width=\textwidth]{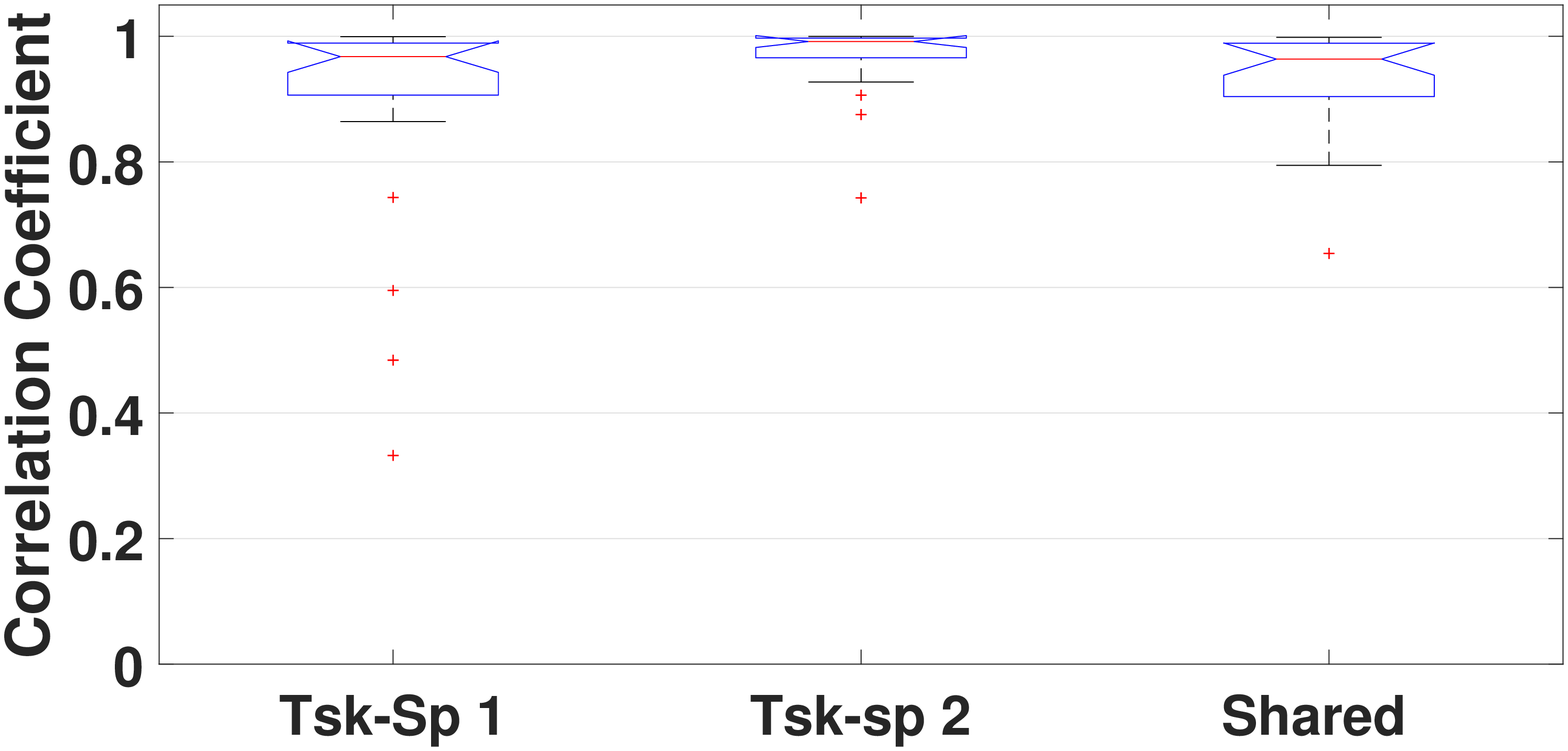}
    \caption{\textit{DoF 1.}}   
    \label{fig:CorrDoF1}
  \end{subfigure}
    \hfill
  \begin{subfigure}[b]{0.32\textwidth}
    \includegraphics[width=\textwidth]{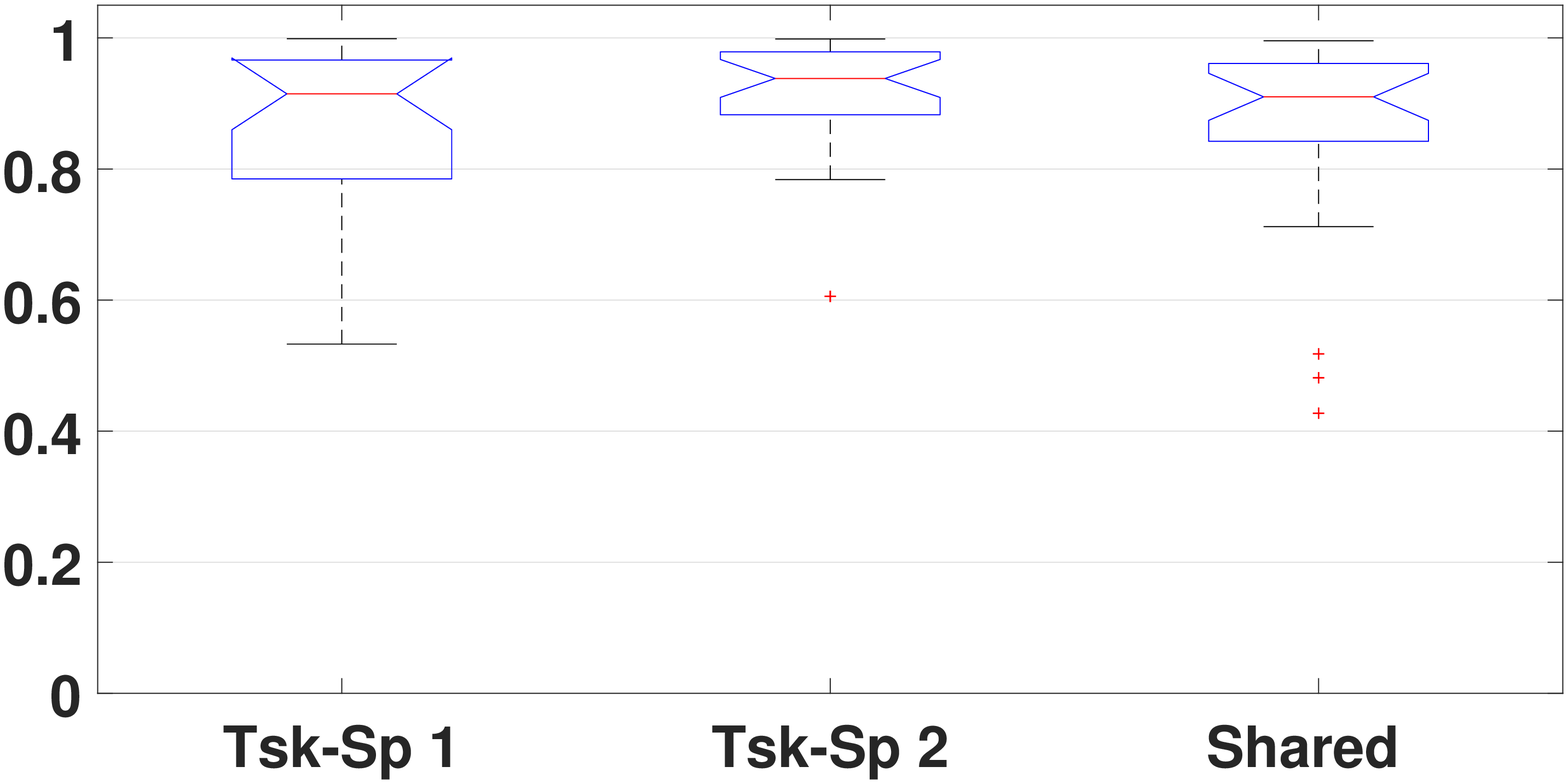}
    \caption{\textit{DoF 2.}}   
    \label{fig:CorrDoF2}
  \end{subfigure}
    \hfill
  \begin{subfigure}[b]{0.32\textwidth}
    \includegraphics[width=\textwidth]{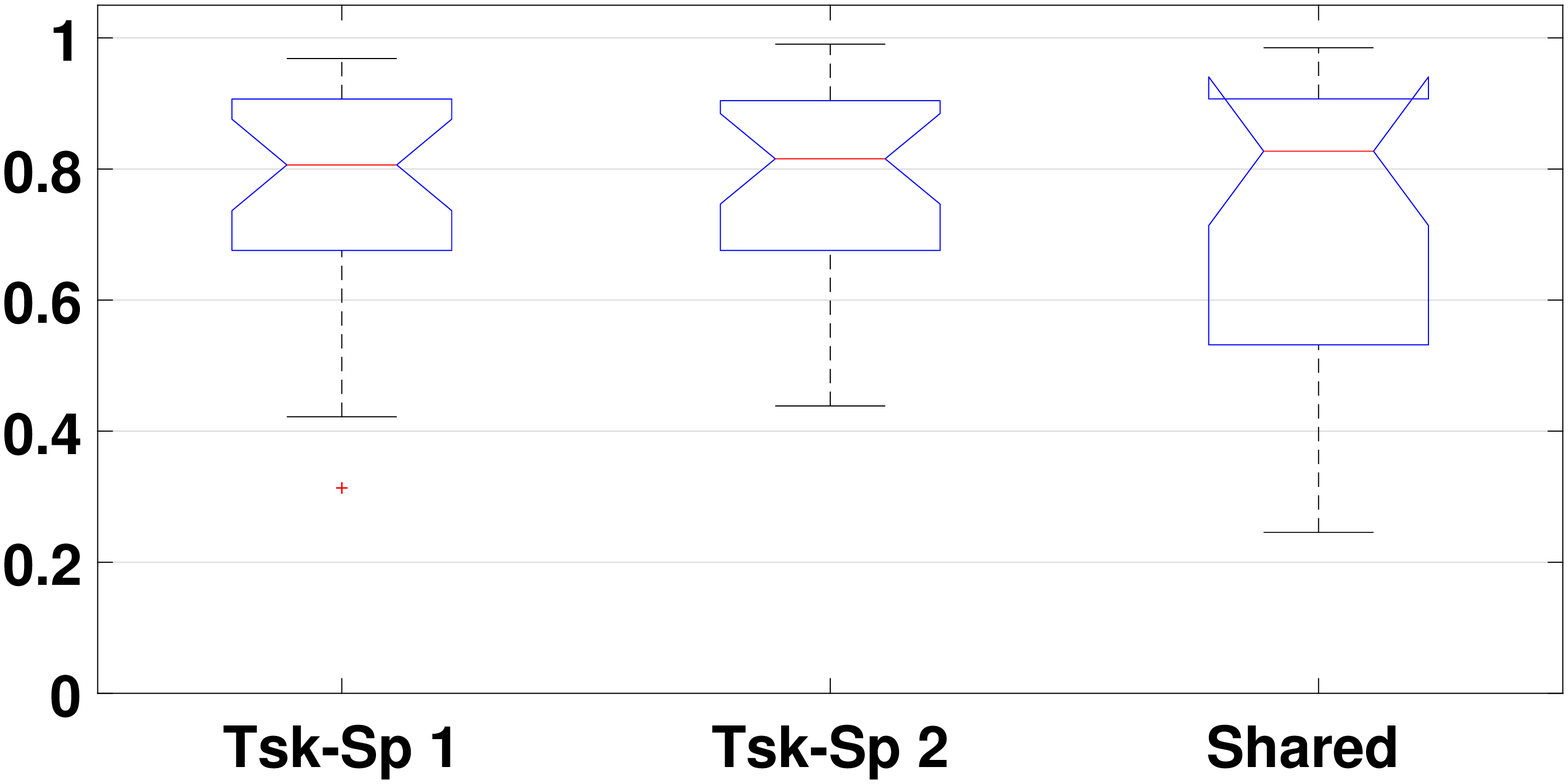}
    \caption{\textit{DoF 3.}}  
    \label{fig:CorrDoF3}
  \end{subfigure}
    
\caption{Boxplots for correlation coefficients between synergies extracted  from the 3 DoFs tensor and from single DoF tensors for the 27 subjects. Each panel shows the correlation coefficients of the 3 synergies (2 task-specific and 1 shared) estimated from DoF1 (\ref{fig:CorrDoF1}), DoF2 (\ref{fig:CorrDoF2}) and DoF3 (\ref{fig:CorrDoF3}) and their respective synergies estimated from the all 3 DoFs tensor. 
  \label{fig:CorrAll}
}
\end{figure*}

The correlation coefficient and cosines of angles were computed between synergies extracted from 1-DoF tensor and their respective synergies estimated by the decomposition of 3-DoF tensors. The mean values of cosine similarity measures were $> 0.88$ as represented in Table \ref{tbl:Cosine}. The correlation coefficients for the 27 subjects is represented as boxplots in Fig.~\ref{fig:CorrAll}. The mean values for DoF 1  were 0.899 and 0.968 for task specific (\textit{tsk-sp}) synergies and 0.936 for the shared synergy as shown in Fig.~\ref{fig:CorrDoF1}. On the other hand, 0.868 and 0.918 were the mean values of the correlation coefficient for DoF 2 task specific synergies and 0.854 for shared synergy (Fig.~\ref{fig:CorrDoF2} ). Finally, 0.777, 0.783 and 0.723 were the mean values for DoF3 as shown in Fig.~\ref{fig:CorrDoF3}.

\begin{table}[b]
\centering
\setlength{\extrarowheight}{0pt}
\addtolength{\extrarowheight}{\aboverulesep}
\addtolength{\extrarowheight}{\belowrulesep}
\setlength{\aboverulesep}{0pt}
\setlength{\belowrulesep}{0pt}
\caption{The mean of cosine angles between synergies extracted from 3-DoFs tensor and single DoF tensors across the 27 subjects.}
\begin{tabular}{c|c|c|c} 
\toprule
\rowcolor[rgb]{0.863,0.863,0.855}      & Task-sp. Synergy 1 & Task-sp. Synergy 2 & ~Shared Synergy  \\ 
\hline
DoF1                                   & ~~~ 0.963               & 0.986                   & ~0.979           \\
\rowcolor[rgb]{0.863,0.863,0.855} DoF2 & ~~~ 0.942               & 0.958                   & 0.957            \\
DoF3                                   & ~~~ 0.881               &  0.887                  & 0.909            \\
\bottomrule
\end{tabular}
\label{tbl:Cosine}
\end{table}

\begin{figure*}[htbp]
	
	\begin{subfigure}[b]{0.32\textwidth}
		\includegraphics[width=\textwidth]{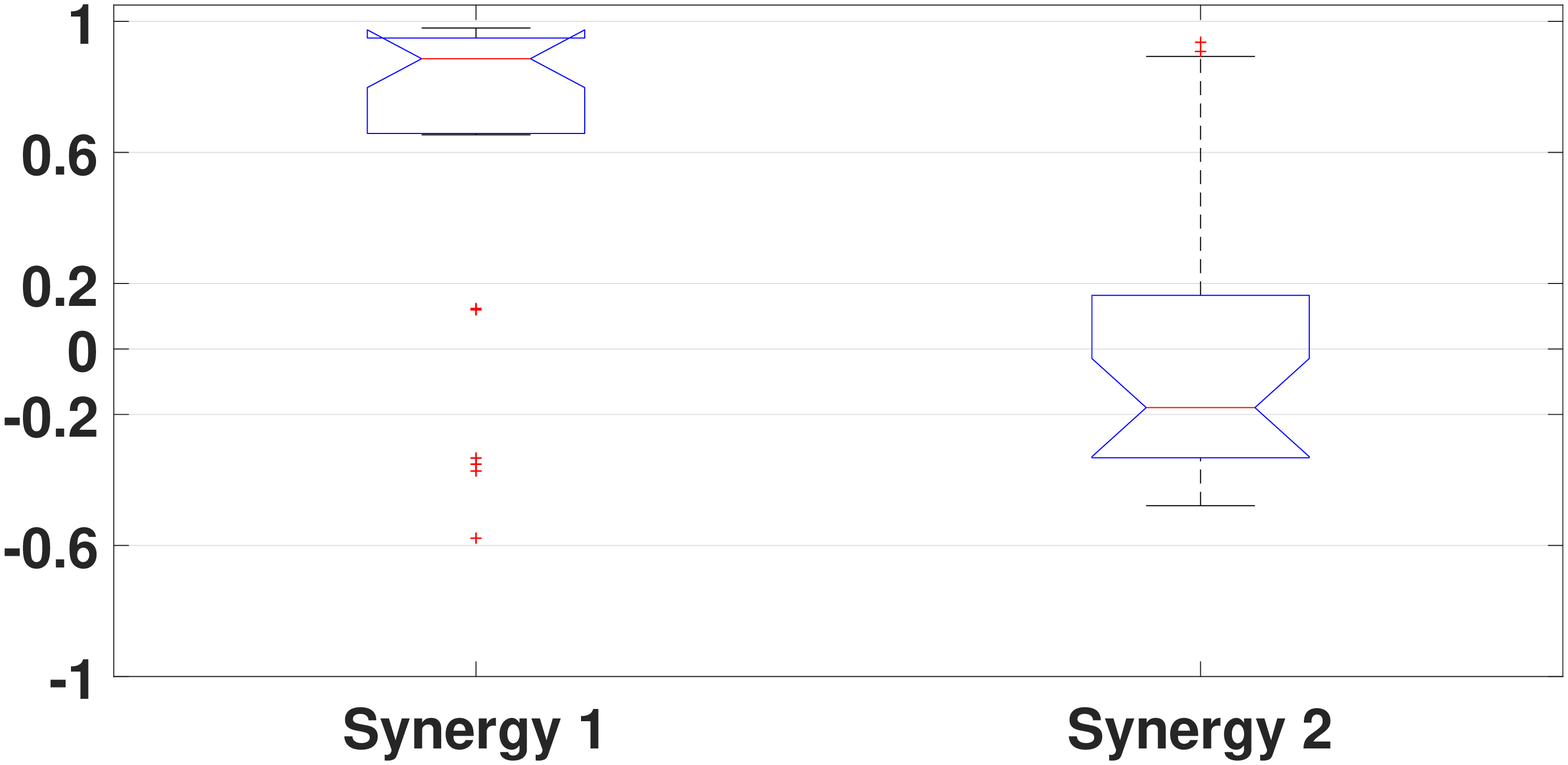}
		\caption{\textit{DoF 1.}}   
		\label{fig_ch6:CorrNMF1}
	\end{subfigure}
	\hfill
	\begin{subfigure}[b]{0.32\textwidth}
		\includegraphics[width=\textwidth]{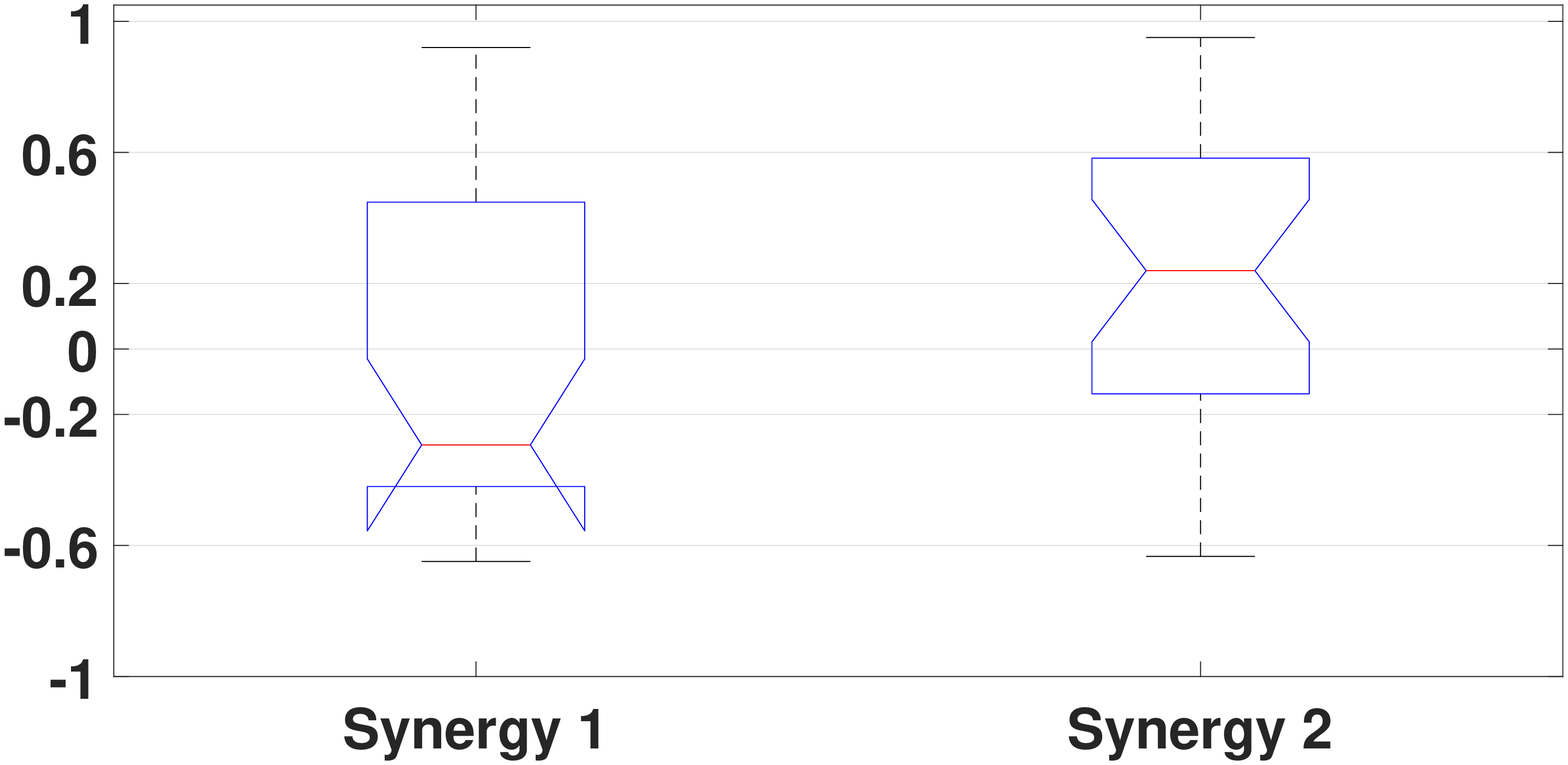}
		\caption{\textit{DoF 2.}}   
		\label{fig_ch6:CorrNMF2}
	\end{subfigure}
	\hfill
	\begin{subfigure}[b]{0.32\textwidth}
		\includegraphics[width=\textwidth]{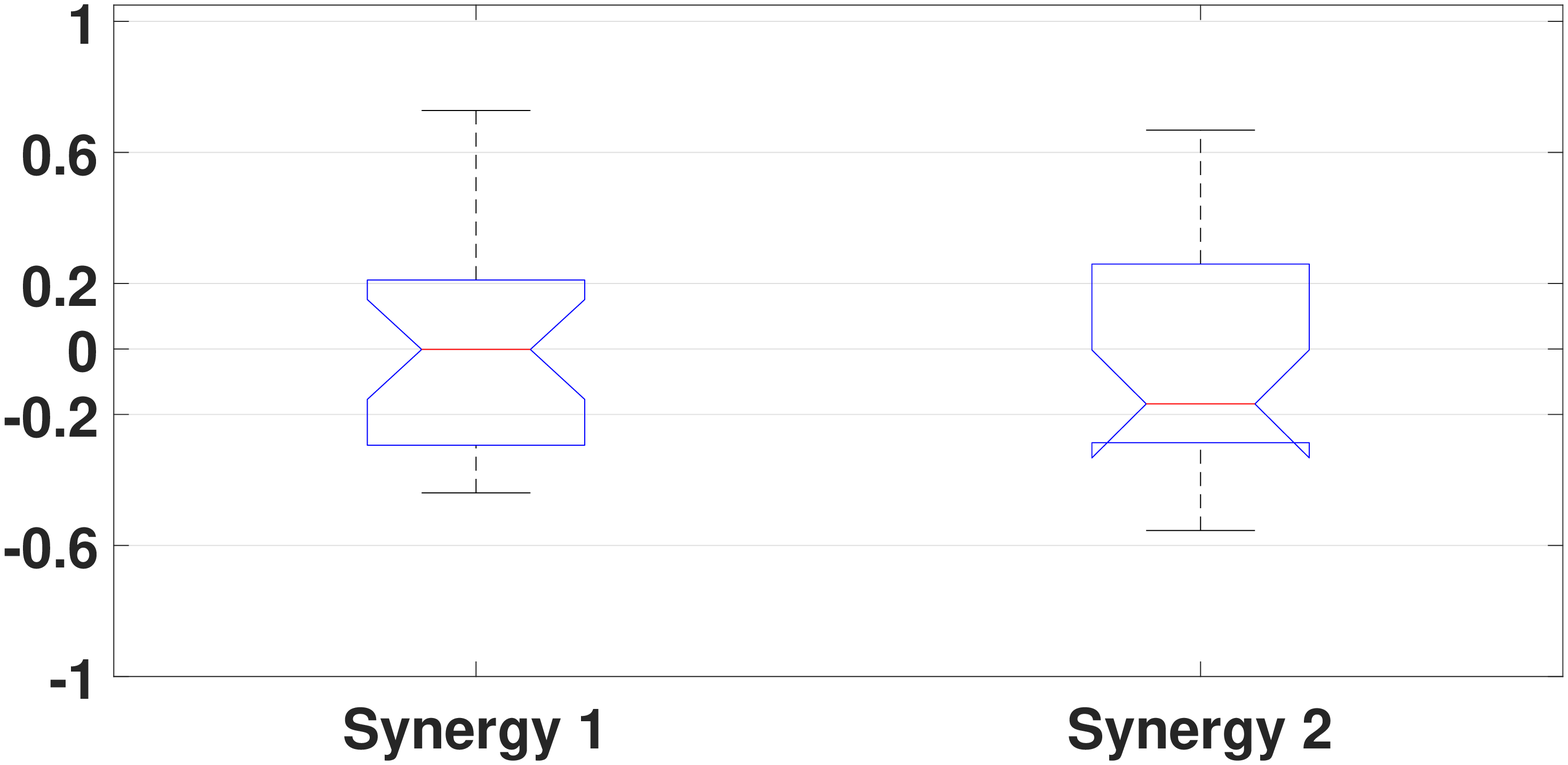}
		\caption{\textit{DoF 3.}}  
		\label{fig_ch6:CorrNMF3}
	\end{subfigure}
	
	\caption{Boxplots for correlation coefficients between the six synergies extracted via NMF from the EMG segment of all 3 DoFs and from single DoF segments for the 27 subjects. 
		\label{fig_ch6:CorrNMFAll}
	}
\end{figure*}

\section{Discussion and Conclusions}

The ability of constrained Tucker decomposition to extract consistent muscle synergies with the increase of task dimensionality from one to three DoFs was investigated to assess the potential use of muscle synergies in proportional myoelectric control. Synergies extracted via constrained Tucker methods from 3-DoF tensor were similar to those extracted separately from 1-DoF tensors as shown in Fig.~\ref{fig:CorrAll} and Table \ref{tbl:Cosine}. This supports the use of tensor factorisation to estimate synergies since the extracted profiles would not depend on the number of DoFs under consideration.

The same approach for comparing synergies extracted from the data of one and three DoFs was applied using NMF instead (plots not included due to space restrictions). The results were notably worse than those of tensor synergies since the NMF approach cannot link the extracted synergies to their respective tasks. Hence, studies utilised muscle synergies for myoelectric control \cite{Jiang2014b,Ma2015a} used to divide the data into a 1-DoF segments and extract 2 task-specific synergies from each segment separately via NMF. Lin \textit{et al.}~\cite{Lin2017} tried to solve this issue using sparse NMF to identify 4 task-specific synergies from 2-DoF segment. However, they need to label extracted synergies since NMF will not extract them in a fixed order. In contrast, the constrained Tucker decomposition approach can estimate consistent identified synergies directly from 3-DoF data. Moreover, the tucker model can extract three synergies for each DoF by incorporating additional shared synergy unlike NMF where only two synergies are estimated for each DoF (one for each task).

To sum up, we explored the potential benefits of  higher-order tensor decomposition for proportional  myoelectric control based on muscle synergies focusing on the ability of the Tucker tensor model to identify consistent muscle synergies from 3-DoFs dataset  directly. Further work is needed to achieve proportional myoelectric control based on synergies computed with tensor decompositions but the results are encouraging.

\bibliographystyle{ieeetr}
\bibliography{BioCAS}

\end{document}